\documentclass[12pt,american,english]{article}
\usepackage[T1]{fontenc}
\usepackage[latin9]{inputenc}
\usepackage[letterpaper]{geometry}
\usepackage{mathtools}
\usepackage{amsthm}
\usepackage{amsmath}
\usepackage{amssymb}
\usepackage{fixltx2e}
\usepackage{graphicx}
\usepackage{setspace}
\usepackage[authoryear]{natbib}
\usepackage{nameref}

\makeatletter
\numberwithin{equation}{section}
\numberwithin{figure}{section}
\theoremstyle{plain}
\newtheorem{thm}{\protect\theoremname}[section]
  \theoremstyle{definition}
  \newtheorem{defn}[thm]{\protect\definitionname}
  \theoremstyle{definition}
  \newtheorem{example}[thm]{\protect\examplename}

\usepackage[all]{xy}
\usepackage{colortbl}
\usepackage{pdflscape}
\usepackage{multirow}
\usepackage{musicography}
\date{}

\usepackage[english]{babel}
\usepackage{hyperref}
\addto\captionsnorsk{\renewcommand{\contentsname}{Contents}}

\makeatother

\usepackage{babel}
  \addto\captionsamerican{\renewcommand{\definitionname}{Definition}}
  \addto\captionsamerican{\renewcommand{\examplename}{Example}}
  \addto\captionsamerican{\renewcommand{\theoremname}{Theorem}}
  \addto\captionsenglish{\renewcommand{\definitionname}{Definition}}
  \addto\captionsenglish{\renewcommand{\examplename}{Example}}
  \addto\captionsenglish{\renewcommand{\theoremname}{Theorem}}
  \providecommand{\definitionname}{Definition}
  \providecommand{\examplename}{Example}
\providecommand{\theoremname}{Theorem}

\begin{document}

\title{Analysis and Visualization of Musical \\
Structure using Networks.}

\author{Alberto Alcal\'a-Alvarez, Pablo Padilla-Longoria\\
Institute for Applied Mathematics (IIMAS), UNAM, Mexico\\
\\
\\
\\}

\maketitle
\begin{abstract}
\thispagestyle{empty}In this article, a framework for defining and
analysing a family of graphs or networks from symbolic music information
is discussed. Such graphs concern different types of elements, such
as pitches, chords and rhythms., and the relations among them, and
are built from quantitative or categorical data contained in digital
music scores. They are helpful in visualizing musical features at
once, thus leading to a computational tool for understanding the general
structural elements of a music fragment.Data obtained from a digital
score undergoes different analytical procedures from graph and network
theory, such as computing their centrality measures and entropy, and
detecting their communities. We analyze pieces of music coming from
different styles, and compare some of our results with conclusions
from traditional music analysis techniques.
\end{abstract}

\section*{Introduction}

In the present work we discuss a framework for analysing symbolic
music information, \emph{i.e.,} considering digital transcriptions
(\emph{e.g.}, MIDI or MusicXML files) of the music fragment or piece
in question, which is parsed (using the Music21 Python library) and
translated as a set of graphs or networks (built using the NetworkX
library in Python) identifying certain musical features of general
interest. This procedure lets us picture and take into account the
overall construction of the piece, by plotting and measuring the occurrence
of different kinds of elements as well as relations among them. Due
to our use of the Music21 Python library, which includes a large
set of tools to deal with digital symbolic music formats such as MIDI
and XML files, for the time being we constrain ourselves to music
written in conventional western notation. That is, into a five line
staff or pentagram, with the usual elements and signs such as measure
bars and rhythmic values of half, quarter, eighth etc. However, this
does not force us to exclusively focus on western academic music,
as the analyzed data may include rhythms expressed as durations in
miliseconds and pitches expressed in cents. This means that, in principle,
we may apply the present framework to microtonal music and unmeasured
music, given that we have it properly codified in a symbolic digital
file. We believe this method may be extended to deal with electronic
textures and continuous sounds in general. 

Several approaches applying graph and network theory for music analysis
have been published in recent years, for example \citet{GrPattMatchPostTon,GrThApprTonMod,ComplMusPatt,TonHarmTopDynNetworks,GrModPopMus}.
Certain aspects in our scope are similar to some of those discussed
in such papers, yet it constitutes a parallel proposal. For example,
we aim to analyse music beyond a particular musical language or genre,
and trying to bring together both vertical and horizontal aspects
of music, with the possibility of including diverse types of musical
objects. Also we incorporate a dynamic approach to the evolution of
families of graphs associated to music data. We focus on the possibilities
of computational analysis and visualization that could be useful for
contemporary concert music and non-Western repertoires, while seeking
consistency with tonal, modal and serial music analysis.

The present article is meant to showcase one possible general working
pipeline for modeling symbolic music information with graphs, and
is in no way an exhaustive. Such graphs may contain elements of very
diverse nature, and for reasons of space, we demonstrate the foundations
of our proposal focusing mainly on pitch information (we consider
a multigraph including duration values).

The present article is broadly structured in the following way: 
\begin{enumerate}
\item In the first section, we develop some of the basic mathematical concepts
and tools; 
\item Next, we describe the general methodology we follow.
\item In the third section we define and illustrate several graphs from
symbolic music data and discuss some of the metrics, properties and
other associated objects we consider for each graph; we also incorporate
the dynamic dimension of our proposal, through the use of sliding
time windows and time series.
\item Finally we discuss the analysis of some graphs associated to music
fragments from three different styles and languages, and present our
conclusions.
\end{enumerate}
\pagebreak{}

\global\long\def\colimrecto{\mathop{\mathrm{col\acute{\imath}m}}}
\renewcommand{\contentsname}{Contents}\tableofcontents{}

\pagebreak{}

\section{Preliminaries.}

\subsection{Graphs and networks.}

A\textbf{ graph} $G$ is a mathematical object consisting of two sets: 
\begin{itemize}
\item A set $V(G)=V$ of vertices or nodes, which in our setting will consist
of musical objects such as pitch classes, chords and rhythms (durations). 
\item A set $E(G)=E$ of edges, which are pairs of vertices, possibly ordered,
in which case the order shows a direction from one vertex to another.
In our setting, edges will represent the concurrence of musical elements,
for example a chord-node will be connected to all pitch-class-nodes
corresponding to the notes it contains, as well as to all the rhythm-nodes
with the duration values the chord takes along the fragment.
\end{itemize}
The \textbf{order} of a graph is its number of vertices, and its \textbf{size}
is its number of edges.

A \textbf{network} is a kind of graph whose edges are all directed
(also called arrows) and elements (both nodes and edges) have an extra
attribute associated to them, which is generically called \textbf{weight}.
In different contexts, this weight may stand for size (when talking
about nodes), capacity or speed (when speaking of arrows) of elements.
The \textbf{degree} of a vertex $v$, $\text{deg}(v)$, is the number
of edges containing $v$, and if edges are directed then we may speak
of the \textbf{in-degree} and \textbf{out-degree} of $v$, respectively
denoted $\text{deg}_{in}(v)\,,\,\text{deg}_{out}(v)$.
\begin{defn}
A \textbf{cluster} in a graph is an induced subgraph with dense inner
connections and sparse interconnections.

The \textbf{density} of a graph is the proportion between the size
of a graph and the maximum possible size of a graph of the same order
(for undirected graphs, $\begin{pmatrix}|V|\\
2
\end{pmatrix}$, the size of the complete graph of order $|V|$).
\end{defn}
In this work, we present graphs whose nodes are either pitch classes,
represented by integers from $0$ to $11$, chords, written as tuples
of pitch classes without repetition, and rhythms, expressed in quarter-length-duration,
'qL' for short. These nodes are tagged as a tuple consisting of a
decimal number together with the legend 'qL'. No more than one edge
between two nodes in our graphs will be allowed. Instead, we will
add a weight to each of them, according to the number of occurrences
of the corresponding relations in the analyzed fragment. We will also
add another weight to the nodes, counting the total duration along
the fragment of the corresponding element in the score, and will not
be considering edges from a node to itself (called loops).

\subsubsection{Node Centrality.}

Vertices in a graph may be considered to be important under different
criteria. One that is particularly useful is to measure\emph{ how
connected} each vertex is to other vertices, that is, how many edges
are incident with that vertex, or how many \emph{neighbors}\textbf{\emph{
}}it has. This leads to defining different measures called \textbf{centralitiy
measures}, which numerically describe the number of connections of
a vertex. When talking about pitches or chords, a node with high centrality
is a pitch or chord that leads to many others, and so may be a pivot
chord or a sort of ``tonic''. On the other hand, nodes with a low
centrality can be taken as passing notes or chords, or perhaps ornamental
elements.

Out of the different notions and definitions of centrality measures,
we consider \textbf{degree centrality} and \textbf{eigenvector centrality}.
\begin{itemize}
\item Degree centrality is the normalised degree of vertices, with respect
to the maximum possible degree of a vertex ($n-1$ in a graph on $n$
vertices). That is, for a vertex $v$ in a graph with $n$ vertices,
its \textbf{degree centrality} is 
\[
\delta(v)=\frac{\text{deg}(v)}{n-1}\,\text{.}
\]

\item \textbf{Eigenvector centrality} takes advantage of the fact that for
non-negative matrices (such as the adjacency matrix of a graph or
network) the Perron-Frobenius Theorem (see theorem \ref{thm:(Teorema-de-Perron-Frobenius)}
in the \nameref{sec:Appendix.}) gives exactly one eigenvector, called
the \textbf{leading eigenvector}, whose entries are all non-negative
(see for example \citet{NetworksNewman}). Moreover, the leading eigenvector
corresponds to the largest eigenvalue of the matrix. Thus, considering
a graph with ordered set of vertices $V=\{v_{1},v_{2},\,...\,,v_{n}\}$,
the \textbf{eigenvector centrality} of $v_{i}$ is the $i\text{\textendash}$th
coordinate of the leading eigenvector of the adjacency matrix of the
graph. That is, it is the $i\text{\textendash}$th coordinate of the
only non-negative solution to the equation
\[
A\mathbf{\bar{x}}=\lambda\mathbf{\bar{x}}\,\text{,}
\]
where $A$ is the adjacency matrix and $\lambda$ its largest eigenvalue.
\end{itemize}

\subsubsection{Entropy.}

Entropy is a term used in different fields to denote some notion or
measure of disorder (randomness) of a system. In Information Theory,
which is the scope we adopt in this exposition, the entropy of a system
is related to Probability Theory, and is inversely proportional to
the ``amount of information'' we can get out of the \emph{message}.
Whereas the information content associated to the event is the logarithm
of the inverse of its probability of occurrence. Equivalently, entropy
is inversely proportional to how easy it is to guess the state of
the system by ``asking yes-or-no-questions'' (see Information Theory). 

Mathematically, entropy is a number which captures at once the weighted
probabilities of occurrence of the elements in the system. As in the
case of communities, there is more than one way to define a numerical
definition of entropy, though all of them behave more or less alike.

One of the usual ways to define the \textbf{entropy} of a set $X=\{x_{1},x_{2},\,...\,,x_{n}\}$
with probability distribution $\{p(x_{1})=p_{1},p(x_{2})=p_{2},\,...\,,p(x_{n})=p_{n}\}$,
where $p(x_{i})$ stands for the probability of occurrence of the
event or element $x_{i}$, is the \textbf{Shannon formula}:
\[
H(X)=-\underset{i=1}{\overset{n}{\sum}}p_{i}\text{log}_{2}(p_{i})\,\text{.}
\]

This number is always bounded below by $0$ and above by $\text{log}_{2}(n)$.
These bounds correspond, respectively, to the case where all probabilites
are $0$ except for one which is equal to $1$ , and the case where
all elements $x_{1},x_{2},\,...\,,x_{n}$ are equiprobable, each with
probability equal to $1/n$.

In the first situation we get 
\[
H(X)=-1\text{�}\text{log}_{2}(1)=-\text{log}_{2}(1)=0
\]
and in the second case: 
\[
H(X)=-\underset{i=1}{\overset{n}{\sum}}\frac{1}{n}\text{log}_{2}(\frac{1}{n})=\underset{i=1}{\overset{n}{\sum}}\frac{1}{n}\text{log}_{2}(n)=\text{log}_{2}(n)\,\text{.}
\]
Such extreme situations may be interpreted, correspondingly, as the
case with most information (when $H(X)=0$, the state of the system
in question is known with absolute certainty) and the one with the
least possible amount of information (when all events or elements
are equally likely to occur, we have less clues about the message
or system).

This statistical definition of entropy has been applied in music analysis,
particularly to the study of harmonic progressions in W.A. Mozart's
work (see \citet{StatAnMusicCorpora}).

\subsubsection{Entropy in networks\label{sub:Entropy-in-networks.}.}

In the case of graphs, there are also several ways to define entropy
(the reader may refer to \citet{EntropyGraphsHist} for a survey on
different entropy measures). One is to associate weights to nodes
and/or edges, and apply the Shannon formula to their distribution.
Such weights may be given by some graph-driven indicator, such as
the number of neighbors or some other centrality measure for nodes.

Another definition worth considering is the \textbf{von Neumann entropy},
which is defined as follows:

Let $A$ be the symmetric weight matrix of an undirected weighted
graph $G$ with vertex set $V=\{v_{1},\,...\,,v_{n}\}$.  Let $D=\text{diag}(d_{1},\,...\,,d_{n})$
where $d_{i}=\text{deg}(v_{i})$. The \textbf{Laplacian matrix of
}$G$ is defined as
\[
L=D-A\text{,}
\]
whose set of eigenvalues $\lambda_{1},\,...\,,\lambda_{n}$ is called
the \textbf{Laplacian spectrum}.

The \textbf{von Neumann entropy} of $G$ is defined as

\[
H_{\text{vN}}(G)=-\underset{i=1}{\overset{n}{\sum}}\frac{\lambda_{i}}{vol(G)}\text{log}(\frac{\lambda_{i}}{vol(G)})\text{,}
\]
where 
\[
vol(G)=\underset{i=1}{\overset{n}{\sum}}\lambda_{i}=tr(L)\text{.}
\]

Thus, the von Neumann entropy of a graph is the Shannon entropy for
the distribution of the eigenvalues of its associated Laplacian matrix.

Given a cluster $K$ in a graph $G$, the probability of a vertex
$v\in K$ to have inner connections in $K$ is given by 
\[
p_{i}(v)\coloneqq\frac{N_{K}(v)}{N(v)}\text{,}
\]
where $N_{K}(v)$ denotes the number of vertices in $K$ which are
neighbors of $v$, and $N(v)$ denotes the total number of neighbors
of $v$ in $G$. 

On the other hand, the probability of $v$ having outer connections
is naturally defined as
\[
p_{o}(v)\coloneqq1-p_{i}(v)\text{.}
\]

Finally, the \textbf{graph entropy} (see, for example \citet{ProteinGraphEntropy})
of a vertex $v\in K$ is given by 
\[
H(v)\coloneqq-p_{i}(v)\log_{2}p_{i}(v)-p_{o}(v)\log_{2}p_{o}(v)\text{,}
\]
and the \textbf{graph entropy of }$G$ by
\[
H(G)\coloneqq\underset{v\in V(G)}{\overset{}{\sum}}H(v)\text{.}
\]

Notice that this quantity does depend on the number of vertices, hence
comparing graphs of different order with this measure might be misleading.
Hence, we will rather work with the average per node, given by:
\[
H(G)\coloneqq\frac{1}{|V(G)|}\underset{v\in V(G)}{\overset{}{\sum}}H(v)\text{.}
\]
With this definition of entropy, we obtain a measure normalised with
respect to the number of nodes in a graph, which reflects its structure
\emph{as a graph}.

\subsubsection{Communities.}

In a graph, communities are subgraphs which are highly connected.
They consist of subsets of elements that have strong relationships
among themselves, and so they outline parts of the graph which are
structurally important. Though there is not a universally accepted
definition for communities in a graph, we can say they are subgraphs
whose vertices are densely connected among themselves, but have few
connections with nodes in other communities. 

There are different algorithms to establish or detect communities
in a graph. In this work we consider the Clauset-Newman-Moore greedy
modularity maximization algorithm (see \citet{CommunitiesGreedyModularity}),
included in the Networkx package (\citet{ExploringNetX}). This algorithm
is based on merging communities that maximise the change in modularity.

\textbf{Modularity} is a statistical measure of the possibility of
finding community structure in a graph. Given a set of communities,
it quantifies the density of connections in each of them, against
the expected number of connections. A higher modularity indicates
there are in fact communities which are very densely connected within,
yet not with other communities. Hence, modularity is inversely proportional
to the density of a graph (a higher density implies a lesser chance
of finding subgraphs with very few connections with other subgraphs).
Formally, it is calculated as
\[
Q(\mbox{\ensuremath{\mathfrak{C}}})=\frac{1}{2|E|}\underset{u,v\in V}{\sum}(A_{u,v}-\frac{\delta_{u}\delta_{v}}{2|E|})\delta_{C_{u},C_{v}}\,\text{,}
\]
where $\mathfrak{C}$ is a set of communities, $A_{u,v}$ is the $(u,v)$-entry
of the adjacency matrix $A$, $\delta_{v}$ is the degree of $v$,
and $\delta_{C_{u},C_{v}}$ is the Kronecker delta for the communities
$C_{u},C_{v}$, the communities of $u$ and $v$, respectively. This
can be also expressed as
\[
Q(\mbox{\ensuremath{\mathfrak{C}}})=\underset{C\in\mbox{\ensuremath{\mathfrak{C}}}}{\sum}\left(\frac{m_{C}}{|E|}-\theta\left(\frac{\delta_{C}}{2|E|}\right)^{2}\right)\text{,}
\]
where $C$ ranges over the communities of the network, $m_{C}$ is
the number of intra-community links, $\delta_{C}$ is the sum of degrees
of nodes in community $C$, and $\theta$ is the resolution parameter,
which establishes a tradeoff between intra-community and inter-community
edges.

Starting with every node as the single member of a community in a
given network, the Clauset-Newman-Moore greedy algorithm iteratively
merges communities seeking to maximise the modularity in every step.
Its outcome is the ``most relevant'' partition of a graph into clusters.
Yields high-modularity communities.

In contrast with the described method for identifying communities
in a network, there is a graph entropy based algorithm (see section
\ref{sub:Entropy-in-networks.}), consisting of clustering nodes together
to minimise entropy. Beginning with a seed cluster, iteratively minimizing
graph entropy yields another clustering algorithm for community detection:
neighbors of each node are added or deleted if such an operation results
in a lower entropy of the cluster. The process stops when it becomes
no longer possible to get a lower entropy in each obtained cluster.
The result is a set of node clusters with minimal entropy.

\subsection{Discrete time series.}

A \textbf{discrete time series }is a sequence of values from observations
made at certain points in time (usually, evenly distributed). The
main attributes for describing time series are:
\begin{itemize}
\item Trends.
\item Seasonality.
\item Presence and behaviour of irregular fluctuations.
\end{itemize}
In this work we consider only finite discrete time series, each containing
a series of measurements of a certain metric of the graphs and networks
associated with a sequence of fragments covering the score under analysis.

Given two discrete time series, possibly of different length, \textbf{dynamic
time warping }(DTW; see, for example \citet{DynTimeWarp}) is a usual
tool for comparing them. DTW\textbf{ }consists of finding an optimal
strictly index-increasing matching of points in both series (in which
each point in a series may be connected to more than one element of
the other). Optimal here means with minimal cost, where the cost is
the sum of all absolute differences between matched points. This yields
a notion of closeness between time series. However, it is not strictly
a distance or metric, since the triangle inequality cannot be guaranteed.

\pagebreak{}

\section{Methodology}

As we mentioned earlier, our proposal seeks, through the use of graphs,
to model and describe diverse music elements and their relations throughout
a given score (more generally, a fragment of a transcription), from
an agnostic data-focused point of view. We relie on the use of computational
tools for materialising our proposal, and hence aim to work with digital
music notation files. It is worth mentioning that the availability
and suitability of such a file involves a whole set of considerations
and difficulties. On the other hand, if no transcription is already
available, it is possible to merely consider a list of chronologically
ordered music events, encoded as some suitable type of digital object. 

The exposition presented here is not meant to be exhaustive, but rather
illustrative of a general procedure applicable to different types
of musical objects and notions. The proposed method is summarised
in the following steps:
\begin{enumerate}
\item Parse data in a digital score. For this purpose we have used the music21
Python library (\citet{Music21Library}), which allows for the use
of several file formats. For the present work we have used MIDI and
MusicXML files.
\item Segment the score in intervals of equal time length (given in number
of bars or seconds). We consider different fixed lengths and frequencies
of such time windows, which seem appropriate for capturing meaningful
features.
\item For each time window, extract objects encoded in the score, such as
pitch classes, vertical events (which we also refer to as chords),
rhythm values, etc.
\item Define graphs and networks containing such objects as nodes, and edges
which represent sequential or simultaneous occurrence of elements.
\item For the resulting graphs, compute several metrics and perform analysis
algorithms: we consider mainly centrality measures, entropies and
community detection. Repeating this for our whole sequence of time
windows along the score yields a family of discrete time series.
\item Plotting and analysis of the resulting time series throughout the
whole score. 
\item Discussion of musical insights which can be drawn from this network-based
analysis, mainly concerning style and form.
\end{enumerate}

\section{Some graphs associated with sequences of musical events.}

For building and analyzing graphs from musical events, we run a Python
script using two main libraries: Music21, to parse and manipulate
digital music notation files, and NetworkX (\citet{ExploringNetX}).
The code we have so far developed also incorporates some standard
libraries for mathematical purposes, such as Numpy and Matplotlib.

After parsing a file in a symbollic digital format (such as .midi,
.xml, .mxl, .abc, etc.), we focus on getting all the chords in a
fragment of the score, together with their durations, and so we get
a list of ordered pairs whose first entry is a tuple of pitch classes
(without repetitions; that is, a pitch class appearing in two or more
different octaves or instruments in the score will appear only once
in the resulting tuple), and whose second entry is a decimal number
which tells us the duration of that chord in quarters ('qL'=quarter
length). We get one of these pairs for each ``event'' in the transcrpition
or score; that is, one chord is added to the list everytime there
is a new sound being registered in any of the staves making up the
score. Next, we shorten that list by deleting all the repetead chords,
adding up the time of the deleted repetitions, and also keeping count
of the number of repetitions of each chord in the whole fragment being
analyzed. In this way, we end up having a list of chords, all different,
together with their cumulative durations and number of ocurrences.
Later we will use these two numbers associated with each chord to
indicate how ``dense'' a chord or pitch class is, by taking a node's
weight to be its total duration, which will be proportional to its
size when plotted, and its number of repetitions as the degree of
transparency of its color (this gives us some information about texture).
We will proceed analogously with the rhtyhms obtained in the original
list (before adding up all durations of each chord), and asigning
each of them a weight, proportional to its total number of occurrences.
In this work we focus on pitch information.

\subsection{Pitch-chord-rhythm (p-c-r) graph.}

After getting this reduced list of chords expressed as tuples of pitch
classes together with their cumulative durations and their respective
number of repetitions, we proceed to build our first graph, which
we call the \textbf{pitch-chord-rhythm (p-c-r) graph}, which encodes
some meaningful relations among pitches of vertical events and the
time intervals between them (the duration of each vertical event,
delimited by the appearance, prolongation or removal of a note). It
is a multipartite graph in which we have three types of nodes: pitch
classes, chords (vectors of pitch classes) and lengths (decimal expression
of duration in quarter note figures). When plotting the p-c-r graph
we identify different types of nodes with different colors: pitch
classes in turquoise, tuples of simultaneous events (which we also
refer to as \emph{chords}) in red, and rhythmic values in blue. We
connect these nodes in the following way:
\begin{itemize}
\item Each chord-node is connected (or adjacent) to all of the pitch-class-nodes
corresponding to its pitch classes, and so every pitch-class-node
is connected to the nodes of all the chords it belongs to.
\item Each chord-node is connected to the rhythm-nodes of all the durations
with which it appears, and so every rhythm-node is connected to the
nodes standing for all the chords played with that rhythm.
\end{itemize}
Note that we could also connect all pitch-class-nodes to rhythm-nodes,
but we leave this out for the sake of clarity, focusing more on durations
and harmony (not melody) in this part of the analysis (later on we
shall incorporate a ``more melodic'' graph). We also consider a
\textbf{p-c-i-r} \textbf{graph}, in which the ``i'' stands for \emph{interval
classes} and interval-nodes are connected to chords (in normal form)
in which consecutive pitches delimit the corresponding interval.

In general, we can consider adding to our graph other kinds of nodes,
coming from any other attribute of musical objects codified in the
parsed file, like dynamics or instrumentation.

Nodes in the multipartite p-c-r graph are assigned weights in two
different ways: by its number of occurrences along the given fragment
(\emph{i.e.}, its frequency), and by the sum of all its duration values
along the fragment. In plots we respectively depict such attributes
as the size of the node and the level of transparency of its color
(see, for example, Figure \ref{fig:p-c-r-graph-JSBach ArtOfFugue-I mm. 1-8}).
In the case of rhythm-nodes we only consider one weight: its number
of occurrences.
\begin{example}
\label{exa: p-c-r example JSBach-ArtOfFugue-I-mm1-8}To exemplify
the p-c-r graph, we take a well-know fragment of tonal classical music:
the first eight measures of J. S. Bach's Contrapunctus I from The
Art of Fugue BWV 1080.
\end{example}
\begin{center}
\begin{figure}
\begin{centering}
\includegraphics[scale=0.5]{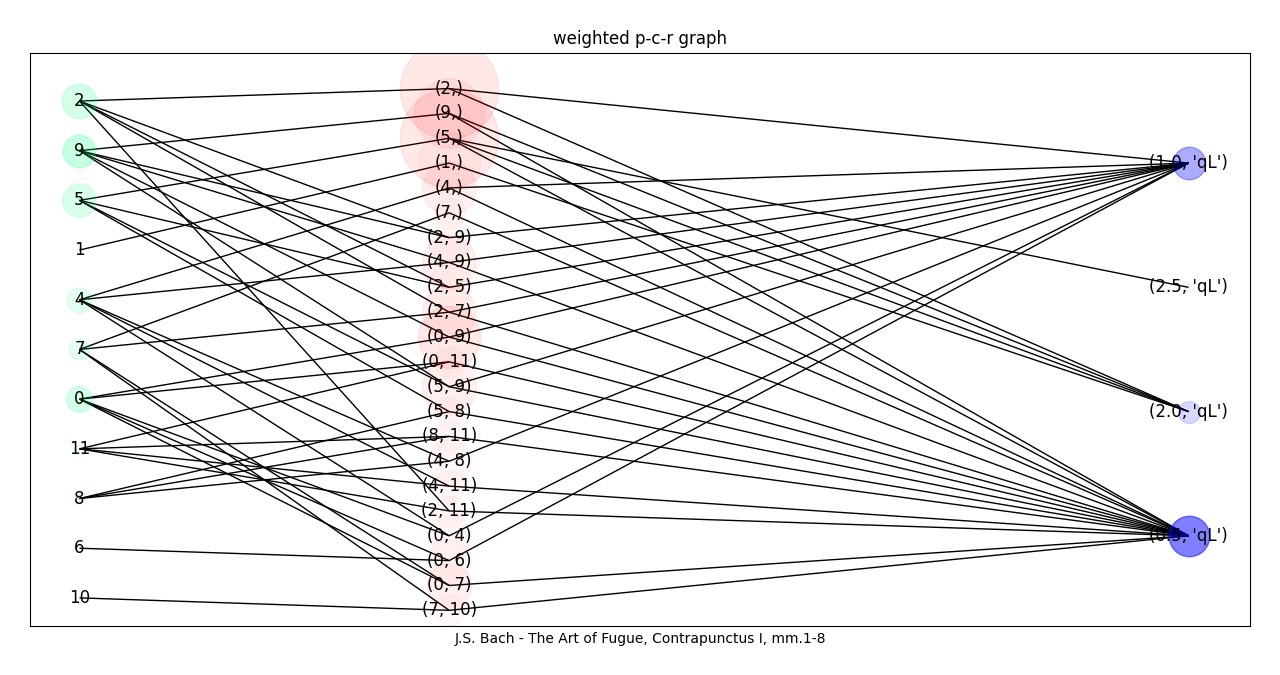}
\par\end{centering}

\protect\caption{\label{fig:p-c-r-graph-JSBach ArtOfFugue-I mm. 1-8}{\footnotesize{}p-c-r
graph for measures $1\text{\textendash}8$ of J. S. Bach's Contrapunctus
I from The Art of Fugue BWV 1080}.}
\end{figure}

\par\end{center}

\subsection{Analysis of the p-c-r graph.}

The metrics considered below for a graph, some of them with very similar
behavior, measure in some way the distribution of information within
the p-c-r graph associated with a music fragment. We consider different
definitions of entropy, as well as centrality measure, density, modularity,
number of communities, and average clustering. In our p-c-r graphs,
this last metric, average clustering, becomes irrelevant (being $=0$
always), since the graph is all connected.

As we saw in Figure \ref{fig:p-c-r-graph-JSBach ArtOfFugue-I mm. 1-8},
taking the p-c-r graph of a whole fragment we get a static map of
all the events (tuples of pitch classes) in the fragment at once,
together with their components (pitch classes and duration in this
case). To incorporate in our analysis the occurrence of events along
a timeline, we consider moving time windows (of different sizes and
regularities) which cover the whole fragment being analised. These
windows are intervals of time of a fixed length which move along the
score, covering the entire fragment, with or without overlapping.
In this paper we consider time windows determined by a fixed number
of bars taken every certain other fixed number of bars. These may
be seen as an interval of bars of a certain radius around a the central
barline of each window (for example, a window comprising mm. 1-8 of
a score may be seen as an interval of radius $4$ bars around barline
$5$.)\footnote{We assume for the time being that the scores we work with are divided
in measures (though the music may not be strictly \emph{metric}),
and work upon subsequences of measures that cover a whole score (which
may be either a complete piece of music, a complete segment or section
of a larger work, or a fragment of a score).}.

We choose a fixed length for our time windows and displace them onward
along the score in question, with a certain regularity, that is moving
their center by a fixed distance at each stage. For each interval
and center in this process we obtain a p-c-r graph, and obtain different
metrics associated with it. This leads us to considering the discrete
time series determined by such numbers. We plot these as polygonal
curves, which together form an ECG of the fragment being analyzed.
Combining ECGs for time windows of different sizes and regularities
we can get a more complete scheme of how information on our musical
parameters is distributed, according to the associated p-c-r graph. 

Of course one immediate issue when running this algorithm is how to
choose the length and regularity of time windows. About this we can
say the following: in order for this technique to be useful, the information
grasped by each window must be neither too much nor too little, whatever
that may mean in each musical context. Windows which are too narrow
result too particular and yield too many points of measurement. On
the other hand, windows which are too large imply all information
is mixed and so measurements are less meaningful. Intuitively, our
sliding window should capture enough information to build meaningful
graphs that yield relevant and distinguishable measurements. Thus,
the length of windows should be compatible with the average length
of phrases, periods or other groupings of music elements in a specific
type of music (for example, windows $4$, $8$ or $16$ bars long
seem reasonable for most music from the Common Practice Period). 

The idea of a fixed length and fixed step for the moving time windows
comes from considering that we may not know in advance how music is
generally structured in a given sample. Also, meaningful groups of
music elements (such as motives, phrases, etc.) are not necessarily
grouped in sequences with the exact same length throughout a piece
or fragment. Ideally, we would have a window englobing each of such
units. This could be the result of fully determining the segmentation
of the score into motives, phrases, etc., or establishing an algorithmic
procedure for varying the lengths and positions of event windows.
In \citet{TonHarmTopDynNetworks}, the author proposes the use of
a change point detection algorithm from signal analysis in order to
segment a score. In a future work we project to develop a segmentation
algorithm proposal based on graph entropy or some other graph descriptor.

\subsubsection{Communities of the p-c-r graph.}

Communities in a p-c-r graph may include different types of nodes:
chords, pitch classes, durations, interval classes, etc. Such music
elements appear to ``work together tightly'', which means they are
consistently coincident. It thus make sense, and has been indeed observed
in samples, that communities of p-c-r graphs associated with tonal
or modal music reflect the stronger harmonic regions and relations.
It is usual to have communities ``leaded'' by a few pitch classes
which group several chords around them. In the tonal context, these
chords are mainly chords containing such pitch classes, together with
passing notes or other ornaments.
\begin{example}
Following example \ref{exa: p-c-r example JSBach-ArtOfFugue-I-mm1-8},
we show the five communities in p-c-r graph from Figure \ref{fig:p-c-r-graph-JSBach ArtOfFugue-I mm. 1-8},
associated with mm. 1$\text{\textendash}$8 of Contrapunctus I from
The Art of Fugue BWV 1080. See Figure \ref{fig:The-five-communities of JSB-ArtOfFugue-I-mm1-8}.

\begin{figure}
\begin{centering}
\begin{minipage}[t]{0.45\textwidth}%
\begin{center}
\includegraphics[scale=0.45]{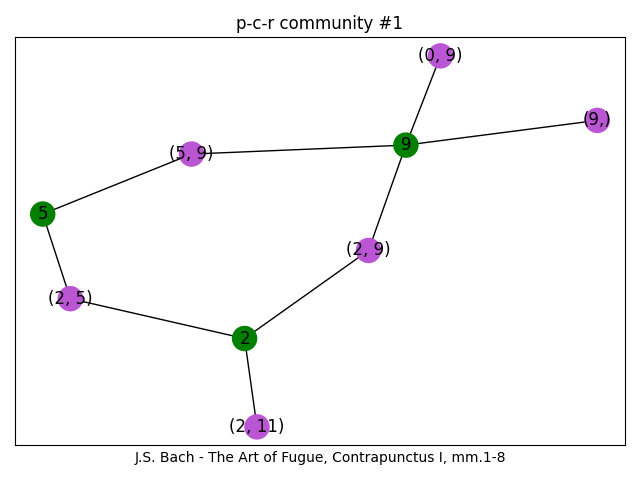}
\par\end{center}%
\end{minipage}%
\begin{minipage}[t]{0.45\textwidth}%
\begin{center}
\includegraphics[scale=0.45]{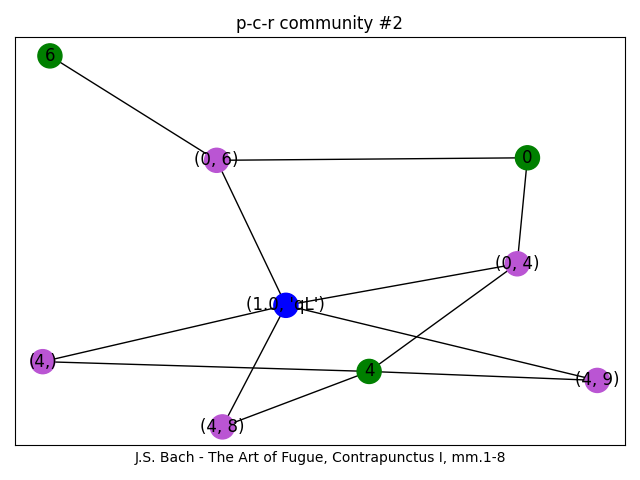}
\par\end{center}%
\end{minipage}
\par\end{centering}

\begin{centering}
\begin{minipage}[t]{0.45\textwidth}%
\begin{center}
\includegraphics[scale=0.45]{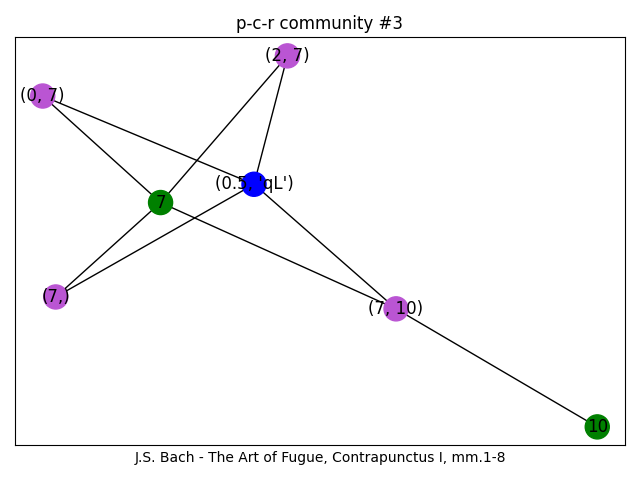}
\par\end{center}%
\end{minipage}%
\begin{minipage}[t]{0.45\textwidth}%
\begin{center}
\includegraphics[scale=0.45]{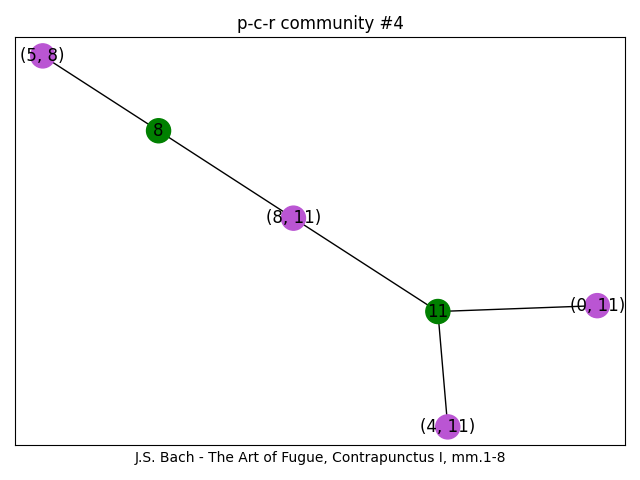}
\par\end{center}%
\end{minipage}
\par\end{centering}

\begin{centering}
\begin{minipage}[t]{0.45\textwidth}%
\begin{center}
\includegraphics[scale=0.45]{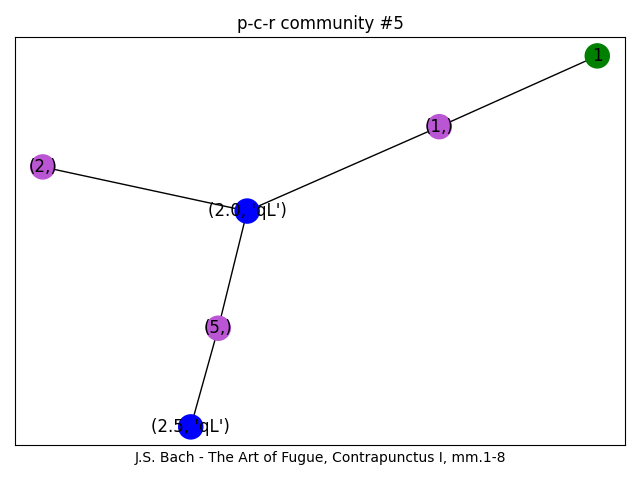}
\par\end{center}%
\end{minipage}
\par\end{centering}

\protect\caption{{\footnotesize{}\label{fig:The-five-communities of JSB-ArtOfFugue-I-mm1-8}The
five communities in the p-c-r graph for mm. $1\text{\textendash}8$
of J. S. Bach's Contrapunctus I from The Art Of Fugue BWV 1080}.}

\end{figure}

The first community of the p-c-r graph clearly corresponds to the
D minor chord (tonic) and some voice leading within and around it.
Community \#2 of this graph shows harmonic constructions around the
third C-E. It includes also the fifth A-E, so it seems to reflect
the A minor triad (minor V chord) and the tritone C-F\sh, which leads
to the subdominant chord (G minor). Besides this, the presence of
the rhythm node tells us that movement involving these chords occurs
mainly in quarter notes. The third community detected by the algorithm
features the pitch-class-nodes $7$ and $10$, corresponding to G
and B\fl, respectively. Looking at the node $(2,7)$ we can relate
this subgraph with the G minor chord (subdominant). Also, we notice
the nodes in this community are highly coincident with the half-quarter
rhythm. Next, community \#4 is easily seen to correspond to the E
Major chord, which is in this case the dominant of the dominant. The
fifth and last of the communities in the p-c-r graph gives us more
information about rhythms of specific pitch classes: D, F and C\sh.
\end{example}

\subsubsection{Entropy of the p-c-r graph.}

For our purpose, we consider different ways of defining the entropy
of a graph, among the many possibilities. All of them tell us how
likely it is to establish a ``predominance'' of certain nodes or
edges (because they measure, more or less broadly, how sparse or condensed
edges and weights in a graph are). Musically we may interpret this
as a measure of the possibility to establish hierarchical relations
among musical objects such as chords, rhythms, intervals or any other
musical parameter to be found codified in the file being analyzed.

If the entropy of a piece of music is low (close to $0$), it is reasonable
that we may establish the particular ``mode'' or ``scale'' it
is in, as well as hierarchies of both chords and rhythms. Moreover,
this same principle may be applied to find out about the relative
importance of sequences of pitches, chords, rhythms, etc. On the other
hand, a higher entropy value means there is a more uniform distribution
of musical elements. Thus, we expect a higher entropy in time windows
which include a noticeable change in texture or harmony, which may
lead to identifying tension/release passages, as well as a potential
segmentation of the score.

We compute the Shannon entropy of pitches, chords and rhythms separately,
taking into account different probability distributions, based on
three different factors: total duration (weight) of nodes, number
of occurrences, degree centrality (the proportion of how many connections
a node has, with respect to the total number of edges in the graph),
and Eigenvector centrality. We also compute the von Neumann entropy
of the resulting graph. As we will see in the examples, several of
these entropy measures share an almost parallel behavior, being most
discordant when calculated for pitch classes. We will discuss this
in depth in the ``Results'' section (section \ref{sec:Results.}).

\subsection{Pitch and chord graphs associated with sequences of events.}

We now discuss some other graphs one may also associate with a music
fragment, which model concurrences and sequences of specific music
elements. As for the p-c-r graph, in all of them the size and degree
of transparency are proportional to the number of occurrences and
total cumulative duration of pitch classes. We keep on exemplifying
the defined graphs with the initial fragment of Contrapunctus I from
The Art of Fugue BWV 1080 (see Example \ref{exa: p-c-r example JSBach-ArtOfFugue-I-mm1-8}).

\subsubsection{Vertical pitch class graph.}

In this graph, nodes are pitch classes, and edges describe vertical
coincidences of notes in the score. That is, two pitch classes are
connected if they appear together in some event from the sequence
considered. The corresponding plot for our basic example is shown
in Figure \ref{fig:Vertical-pitch-class JSB-ArtOfFugue-I-mm1-8}.

\begin{center}
\begin{figure}
\begin{centering}
\includegraphics[scale=0.65]{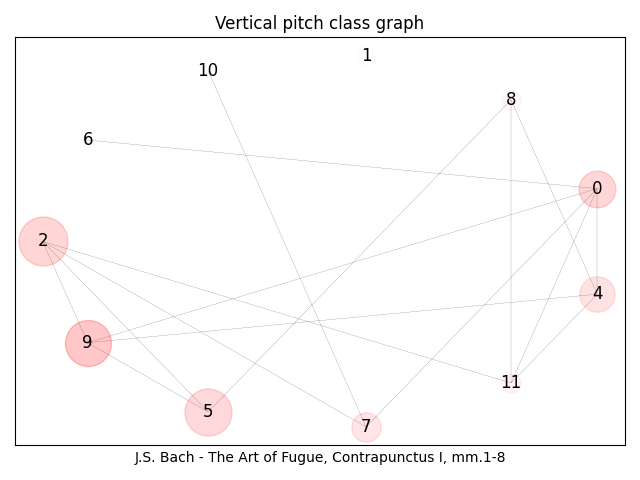}
\par\end{centering}

\protect\caption{{\footnotesize{}\label{fig:Vertical-pitch-class JSB-ArtOfFugue-I-mm1-8}Vertical
pitch class graph for mm. $1\text{\textendash}8$ of J. S. Bach's
Contrapunctus I from The Art Of Fugue BWV 1080}.}

\end{figure}

\par\end{center}

\subsubsection{Horizontal pitch class graph.}

A directed graph whose nodes are picth classes, in which arrows represent
sequence in time. Given a sequence of vertical events there is one
arrow from each pitch class in an event to every pitch class contained
in the next event (see Figure \ref{fig:Horizontal-pitch-class JSB-ArtOfFugue-I-mm1-8}.).
Thus, paths in this network do not strictly correspond to melodies
or harmonic/contrapuntal voices, but rather show pitches as elements
of events which belong to events followed by events containing certain
other pitches. 

This model for horizontal pitch relations does not assume any particular
voice leading scenario. Two consecutive events may contain different
numbers of pitch classes, and in passing from one to another we might
encounter voice crossings, unexpected leaps, or a melody ``jumping''
between registers and/or staves (\emph{e.g.}, timbre melody, named\emph{
klangfarbenmelodie} in the 2nd. Vienese School). In order to analyze
voice leading computationally, it is necessary to establish how to
locate and define melodic movement, which is not univocally defined
in general. For example, keeping track of voice-leading in a post-tonal
setting can be tricky. Hence the choice of not assuming any general
voice-leading criteria, and considering every possible horizontal
relations among pitches from consecutive events.

\begin{center}
\begin{figure}
\begin{centering}
\includegraphics[scale=0.65]{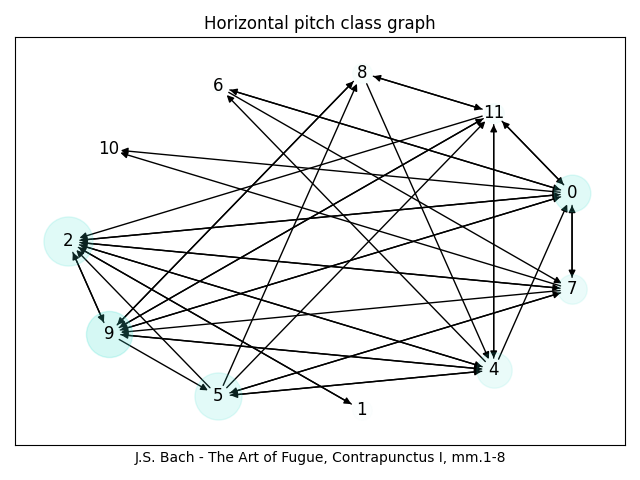}
\par\end{centering}

\protect\caption{{\footnotesize{}\label{fig:Horizontal-pitch-class JSB-ArtOfFugue-I-mm1-8}Horizontal
pitch class graph for mm. $1\text{\textendash}8$ of J. S. Bach's
Contrapunctus I from The Art Of Fugue BWV 1080}.}
\end{figure}

\par\end{center}

\subsubsection{Event or chord sequence graph.}

In this network we merely represent the sequence of vertical events
(\emph{chords}), considering a weight of the arrow, given by the number
of occurrences of the corresponding event progression. Cycles in this
graph represent harmonic cycles which are \emph{feasible} according
to chord progressions score. That is to say, they can be considered
as valid harmonic paths. It is not always the case that these cycles
appear as an actual progression in the score, since it may happen
that such chord connections do not follow each other sequentially
in the score. Nevertheless, we can grasp this by looking at smaller
time windows and keeping track of how cycles arise. The chord graph
for our example is depicted in Figure \ref{fig:Event-sequence-graph JSB-ArtOfFugue-I-mm1-8}.
This graph is considered also in \citet{TonHarmTopDynNetworks}, where
it is then considered itself as a time series.

\begin{center}
\begin{figure}
\begin{centering}
\includegraphics[scale=0.65]{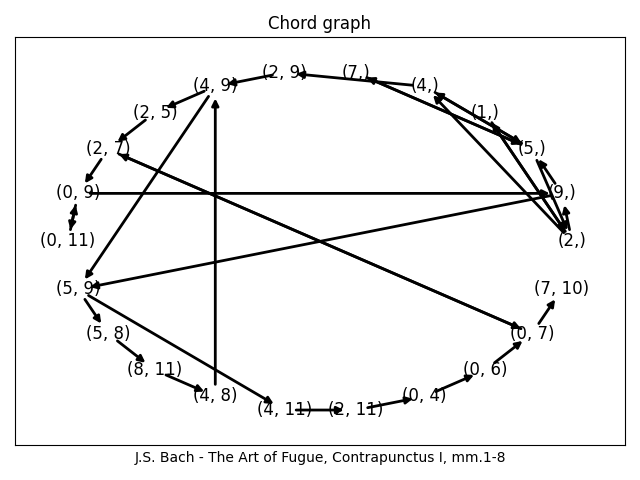}
\par\end{centering}

\protect\caption{{\footnotesize{}\label{fig:Event-sequence-graph JSB-ArtOfFugue-I-mm1-8}Event
sequence graph for mm. $1\text{\textendash}8$ of J. S. Bach's Contrapunctus
I from The Art Of Fugue BWV 1080}.}
\end{figure}

\par\end{center}

These are just examples of graphs and networks which can be universally
associated with a fragment of any music score, each modeling relations
within a specific aspect of music. Further, we may consider graphs
encoding the relations between changes in rhythm, instrumentation,
dynamics or texture, which we expect exploit, in future work, under
the same approach here exposed.

\subsection{Time series and ECG plots from metrics of associated networks.}

\begin{figure}
\begin{centering}
\includegraphics[scale=0.5]{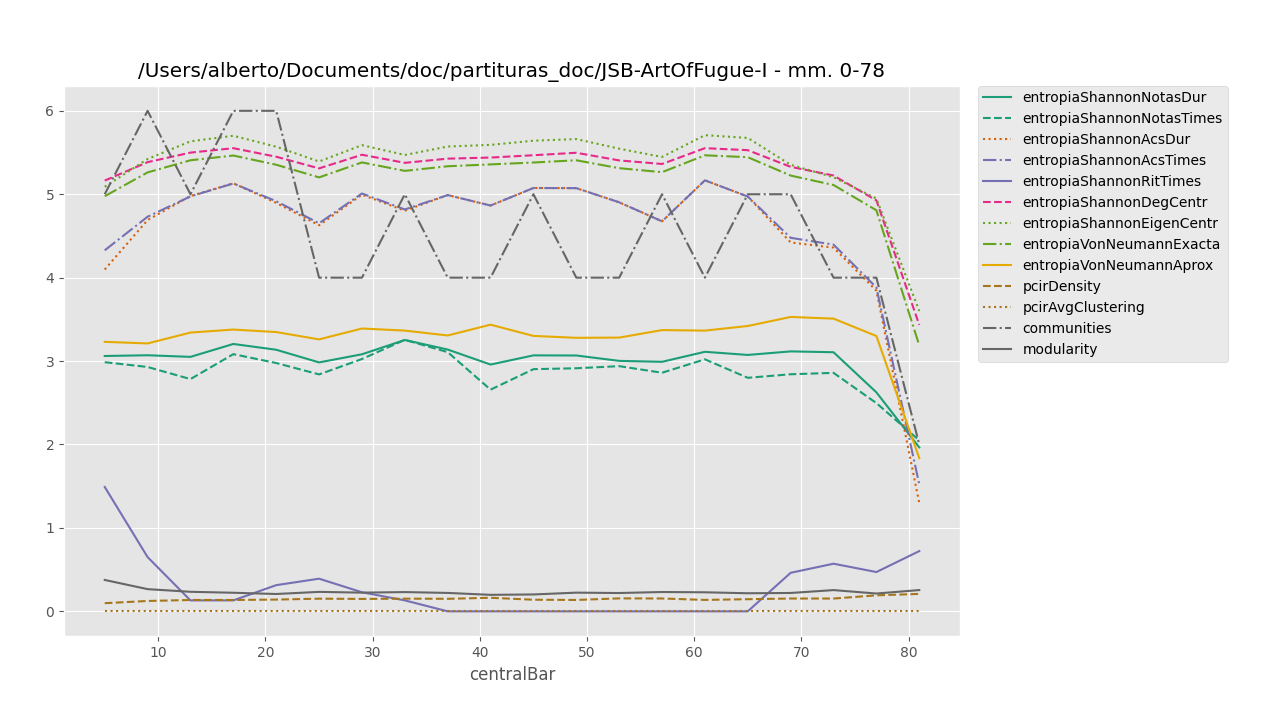}
\par\end{centering}

\protect\caption{{\footnotesize{}\label{fig:ECG-JSB-ArtOfFugue-I_prueba_8cc_cada_4}ECGs
for metrics from the p-c-r graph for mm. $1\text{\textendash}8$ of
J. S. Bach's Contrapunctus I from The Art Of Fugue BWV 1080, taking
time windows of length $8$ mm., every $4$ mm.}.}
\end{figure}

For each length and frequency of our moving time windows, the values
for each metric (centralities, entropies, etc.) of a graph associated
to the corresponding music fragment yield a discrete time series.
Given a length and frequency for time windows covering a score, plotting
together all of these time series for the p-c-r graphs associated
with the given music fragment, results in an ECG type of plot (see
Figures \ref{fig:ECG-JSB-ArtOfFugue-I_prueba_1c_sin_empalmar}$\text{\textendash}$\ref{fig:ECG-JSB-ArtOfFugue-I_prueba_8cc_cada_4}).
As we said before, a common way of comparing time series is dynamic
time warping, which we will apply to several samples in the next section.

Following our example from The Art of Fugue (Contrapunctus I) we briefly
review the ECG type plots obtained from time series of several metrics,
for different time window sizes and frequencies. The metrics computed
here are:
\begin{enumerate}
\item Shannon entropy of notes according to their accumulated duration throughout
the fragment.
\item Shannon entropy of notes according to their number of occurrences
in the fragment.
\item Shannon entropy of chords (pitch class vectors) according to their
accumulated duration throughout the fragment.
\item Shannon entropy of chords according to their number of occurrences
in the fragment.
\item Shannon entropy of duration values according to their number of occurrences
in the fragment.
\item Shannon entropy of the p-c-i-r graph according to the degree centrality
of nodes.
\item Shannon entropy of the p-c-i-r graph according to the eigenvector
centrality of nodes.
\item Von Neumann entropy of the p-c-i-r graph.
\item Density of the p-c-i-r graph.
\item Number of communities of the p-c-i-r graph.
\item Modularity of the p-c-i-r graph.
\end{enumerate}
The first five entropies computed are not graph-specific, but rather
the Shannon entropy of the distributions of durations and frequency
of elements, which are viewed as weights of nodes in the graphs we
have described. The rest of descriptors describe graph-specific properties
of the p-c-i-r graph. This lets us keep track of the evolution of
entropy for each music element considered (pitch classes, pitch class
vectors and duration values), while looking at how the entropy of
the whole multipartite graph as well as its communities, density and
modularity change.

ECG plots can be interpreted as a map of the whole score that shows
the change in variety of several types of elements (in this case,
duration values of events, chords, pitch classes, communities, etc.). 
\begin{example}
\begin{figure}
\begin{centering}
\includegraphics[scale=0.5]{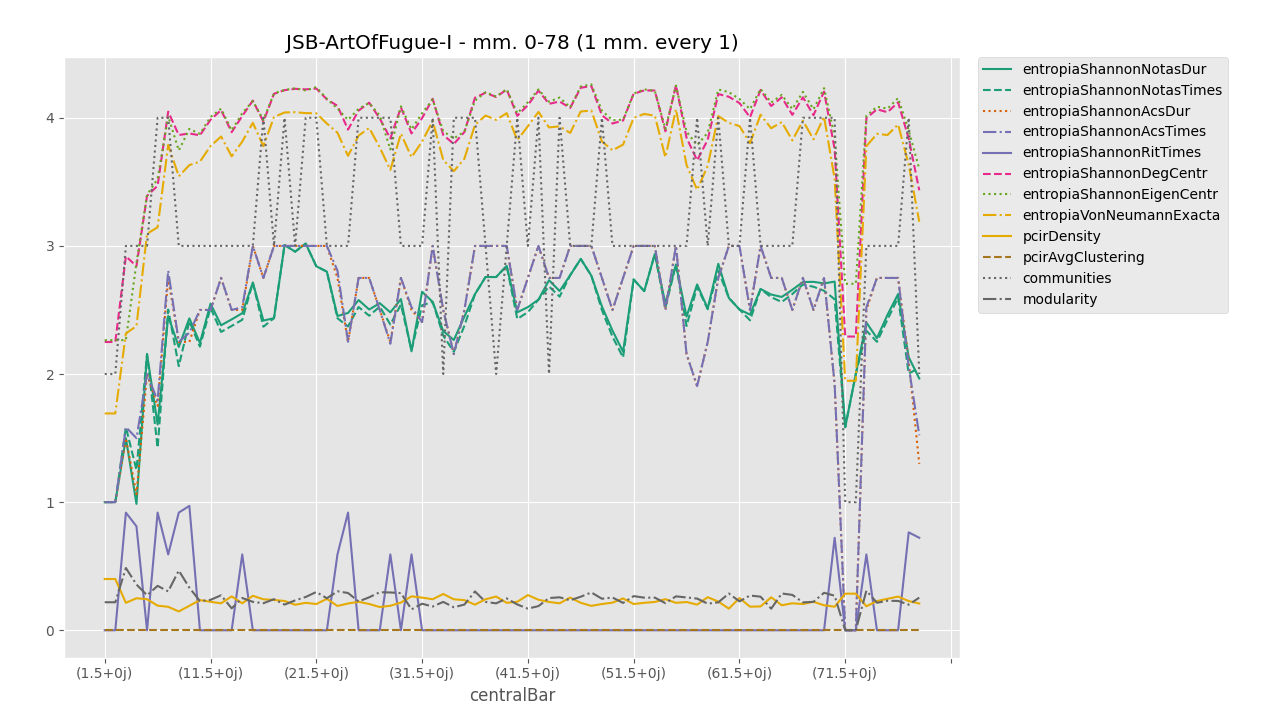}
\par\end{centering}

\protect\caption{{\footnotesize{}\label{fig:ECG-JSB-ArtOfFugue-I_prueba_1c_sin_empalmar}ECGs
for metrics from the p-c-r graph for mm. $1\text{\textendash}8$ of
J. S. Bach's Contrapunctus I from The Art Of Fugue BWV 1080, taking
time windows of length $1$ measure, without overlapping}.}
\end{figure}

\begin{figure}
\begin{centering}
\includegraphics[scale=0.5]{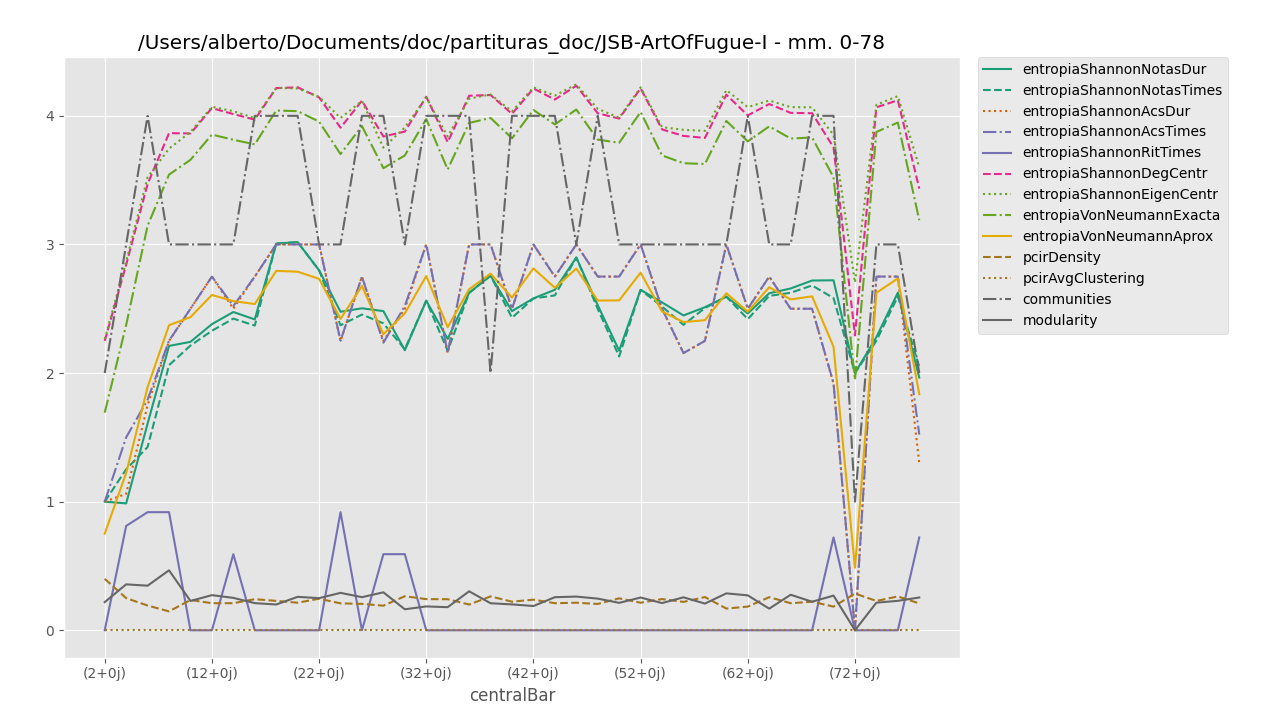}
\par\end{centering}

\begin{centering}

\par\end{centering}

\protect\caption{{\footnotesize{}\label{fig:ECG-JSB-ArtOfFugue-I_prueba_2cc_sin_empalmar}ECGs
for metrics from the p-c-r graph for mm. $1\text{\textendash}8$ of
J. S. Bach's Contrapunctus I from The Art Of Fugue BWV 1080, taking
time windows of length $2$ mm., without overlapping}.}
\end{figure}

\end{example}
\begin{figure}
\begin{centering}
\includegraphics[scale=0.5]{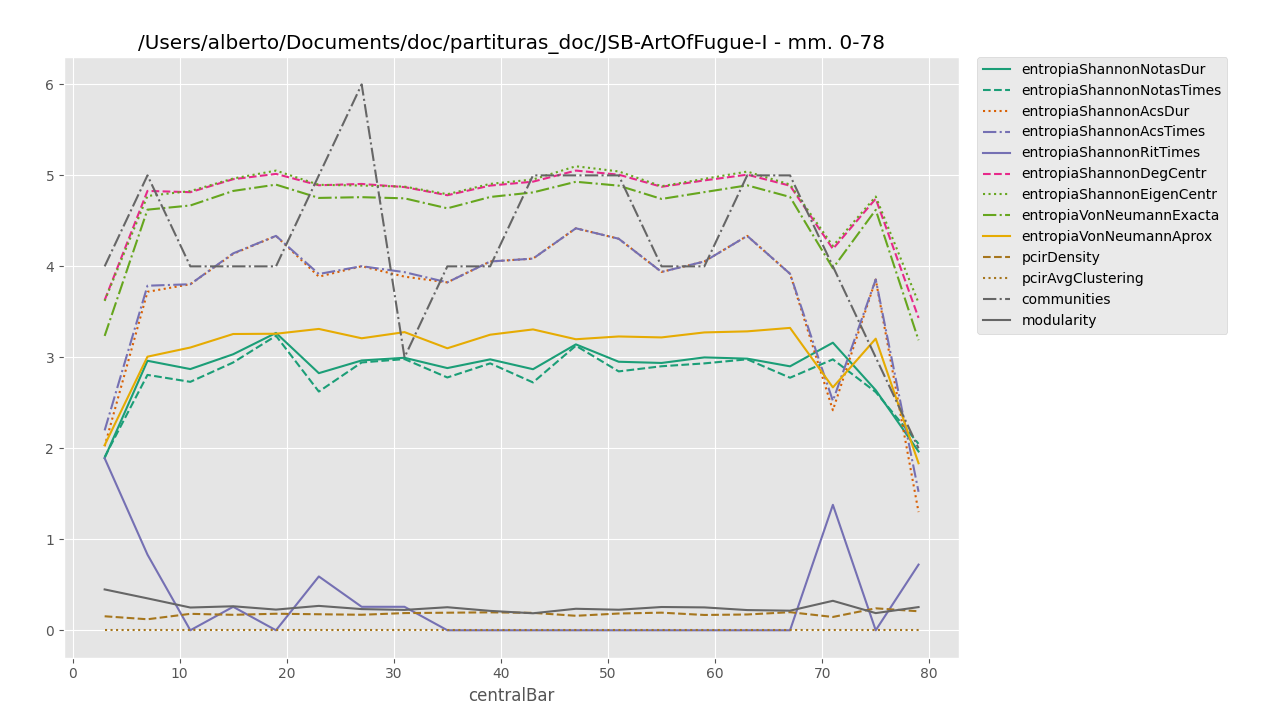}
\par\end{centering}

\protect\caption{{\footnotesize{}\label{fig:ECG-JSB-ArtOfFugue-I_prueba_4cc_sin_empalmar}ECGs
for metrics from the p-c-r graph for mm. $1\text{\textendash}8$ of
J. S. Bach's Contrapunctus I from The Art Of Fugue BWV 1080, taking
time windows of length $4$ mm., without overlapping}.}
\end{figure}
\begin{figure}
\begin{centering}
\includegraphics[scale=0.5]{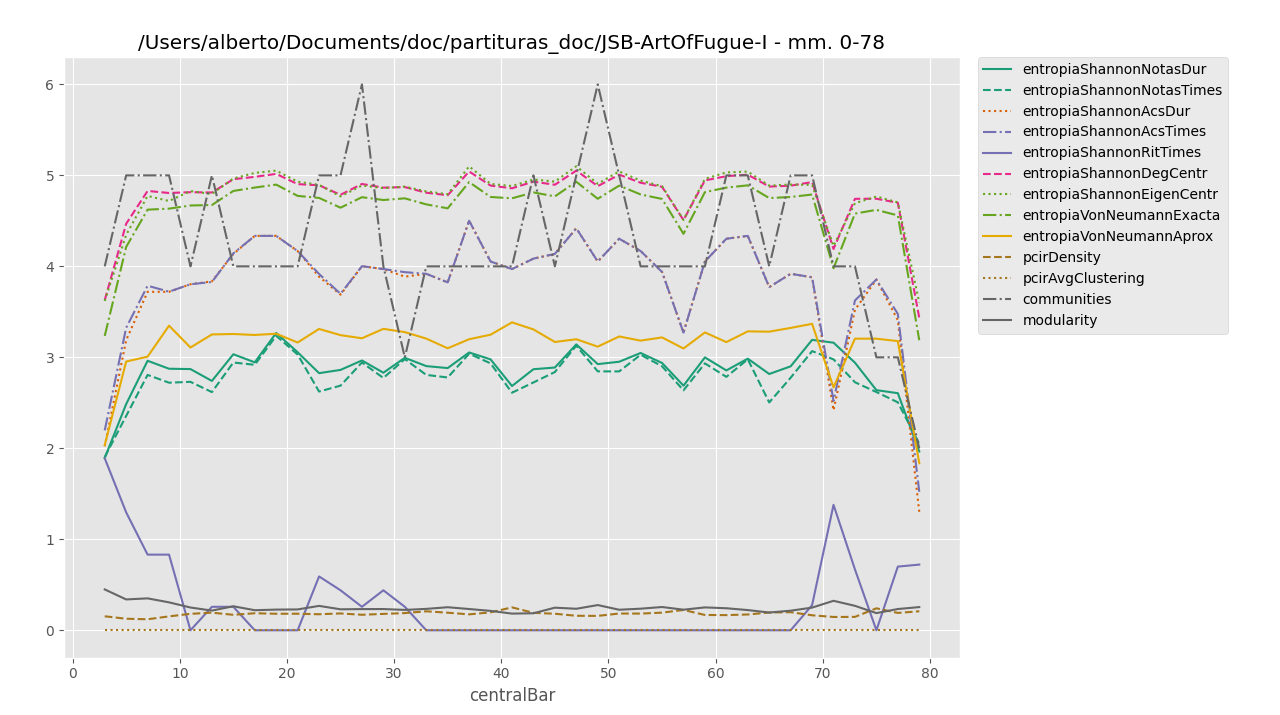}
\par\end{centering}

\protect\caption{{\footnotesize{}\label{fig:ECG-JSB-ArtOfFugue-I_prueba_4cc_cada_2}ECGs
for metrics from the p-c-r graph for mm. $1\text{\textendash}8$ of
J. S. Bach's Contrapunctus I from The Art Of Fugue BWV 1080, taking
time windows of length $4$ mm., every $2$ mm.}.}
\end{figure}
We can make the following general remarks about all of the above plots:
\begin{itemize}
\item The values of entropy calculated with respect to degree centrality
and eigenvector centrality are very close to the von Neumann entropy
of the p-c-r graph, usually with the former being bounded by the other
two. 
\item Shannon entropy of chords according to their number of occurrences
is almost equal to entropy with respect to their total duration.
\item The five entropy metrics mentioned in the first two points above are
almost parallel to each other.
\item As to the values of Shannon entropy of pitch classes with respect
to occurrences and duration, they are near each other and show to
have very similar behavior, they differ more than the groups of entropy
computations discussed above.
\item Entropy of rhythms (taking their frequency distribution) follows a
very different contour, sometimes contrary and in occasions independent
from other metrics. It shows a long flat region with value $0$, corresponding
to a long section of the score in which events occur at even intervals
of time, resulting in only one rhythmic value (therefore their entropy
equals $0$ throughout those bars).
\item There are some evident common peaks or valleys in the ECGs for several
metrics at a certain time or around a certain time. In the next section
we will see these often coincide with section changes.
\end{itemize}
 The fact that different entropy measures have a very similar behaviour
may lead to keeping only a few of them. Nevertheless, the slight crosses
observed between some of them could be meaningful for style classification.
This issue will be addressed in future papers.

Besides this general considerations, we are interested in contrasting
plots corresponding to different time window settings, and what can
that analysis tell us about the music. This involves translating plots.
ECGs can be regarded as a general map of how information in the score
changes, from either a statistical and graph-theoretical point of
view. That is, how much variety or monotonicity of elements is found
in each time window. Thus, when looking at this plots for a whole
score, rather than looking at values of music elements themselves,
it is the change in their distribution and relations.

\section{\label{sec:Results.}Results.}

We begin analyzing a fragment taken from a well-known example of European
classical music: the first counterpoint of J.S. Bach's 'The Art of
Fuge'. We get its principal attributes and show its associated graphs.

J.S. Bach - The Art of Fugue, Contrapunctus I, mm.1-8.

\#eventos: 34

\#eventos/totalDurAcs= 32.0 da numero promedio de eventos por cuarto

totalDurNotas= 60.5/totalDur= 32.0 da numero promedio de voces por
cuarto

totalTimesNotas= 100/totalDur= 32.0 da densidad promedio por cuarto

Entrop�a de Shannon de clases de altura c.r. a la duraci�n: 2.999988562883508 

Entrop�a de Shannon de clases de altura c.r. a las repeticiones: 2.9564684817727676

Entrop�a de Shannon de Acordes c.r. a la duraci�n: 4.045816369049408 

Entrop�a de Shannon de Acordes c.r. a las repeticiones: 4.3149727675300324

Entrop�a de Shannon de Ritmos (c.r. a las repeticiones): 1.5830435300531966

Entrop�a de Shannon de la gr�fica c.r. a la centralidad de grado:
4.969583379448451

Densidad de la gr�fica de alturas-acordes-ritmos: 0.1036036036036036

Aglomeraci�n promedio (average clustering) de acordes: 0.0 \# comunidades:
5 

Aglomeraci�n promedio (average clustering) de clases de alturas horizontales:
0.5536966149182225 Densidad de la gr�fica de clases de alturas horizontales:
0.4727272727272727

Aglomeraci�n promedio (average clustering) de acordes: 0.0256993006993007
Densidad acordes: 0.07142857142857142 

Tama�o de una base de ciclos de acordes: 6

~

Luna: Bach (cc. 1-16), Bartok (cc. 1-19), Brubeck (cc. 1-15) y Haydn
(cc. 1-18). 

vs

Bach (art of fugue, Contrapunctus I, cc.1-8, 8-16cc., 1-22), 

de 8 en 8 cc., cada 4: 

cc.1-8: 5 coms.
\begin{itemize}
\item Com. 1: la-do-mi ; tercera mayor: do-mi, mi-sol\#, fa-la; negra
\item Com. 2: sol\#-si-fa ; tercera menor (segunda menor): fa-sol\#, re-fa,
sol\#-si, si-re ; 
\item Com. 3: sol-sib ; cuarta justa: si-mi, re-sol, sol-do ; corchea
\item Com. 4: do\#-re ; re-la, fa ; blanca
\item Com. 5: fa\#; cuarta aum: do-fa\# 
\end{itemize}
cc.5-12: 6 coms.
\begin{itemize}
\item Com. 1: do-fa-(fa\#) ; (2a. m) do-mi-fa, si-do , (4a. A) do-fa\# ;
do-fa-sol; 
\item Com. 2: re-sib ; 3a. m: re-fa-la, re-fa, sol-sib, si-re ; SibM, sib-re,
re-sol ; corchea
\item Com. 3: mi-la ; 4a. J mi-la, si-mi, re-mi-la, re-sol-la , (2a. M)
\item Com. 4: (sol\#) ; 3a. M mi-sol\#, fa-la, LaM, do-mi ; negra
\item Com. 5: do\#-sol ; LaM7, do\# dim ,sol dim 
\item Com. 6: si; SolM7 sol\#-si
\end{itemize}
cc.13-20: 6 coms.
\begin{itemize}
\item Com. 1: sol\#-si-fa ; sol\# dim FaM, SolM7, MiM, re-fa, sim ; corchea
\item Com. 2: sol-sib ; 3a. m: Solm, SolM, Lam7m, Mim, DoM, mi dim; 
\item Com. 3: do\# ; 3a. M, 2a. m: LaM7, MiM (FaM, SibM) ; (negra)
\item Com. 4: re-la; 4a. J, re m, re-la, mi-la
\item Com. 5: mi; 2a. M la m, MiM7 , si dim, DoM7 
\item Com. 6: do-fa\# ; ReM7, fa\# dim
\end{itemize}
The set of graphs associated with a music fragment discussed in this
text yield both a quantifiable as well as a visual tool for music
analysis in a general musical context.

Also, the present work points toward a possible definition of \emph{musical
meaning} from the point of view of network and time series analysis
and visualization: we may consider each type of music element together
with its attributes as an element of different networks, which includes,
for example, its neighbors, the paths or cycles from/to/through it,
as well as their attributes, and how these change along a certain
lapse of time or interval of events.

From the plots of graphs associated to a sequence of vertical events
we can trace some principal elements of Schenkerian analysis (the
reader may refer to \citet{SchenkerianForte}): 
\begin{itemize}
\item Looking at how an associated graph mutates along the fragment, and
filtering by weight we can recover certain elements of the different
levels of deepness of musical structure (again, surface/subjacent
level).
\item Form determined by clear changes in texture: apparently the analysis
of the ECGs obtained can help track certain changes in form; smaller/bigger
time windows -> surface/subjacent level.
\end{itemize}
On the other hand, from the analysis of ECG plots for moving time
windows along a fragment we obtain results relatable to both Schenkerian
analysis and the generative analysis perspective (\citet{LerdahlJackendoffGTTM}).
In a way, the obtained ECGs share certain features with grouping,
time-span reduction and prolongational reduction, as defined in the
Generative Theory of Music: we are considering a fixed interval of
time (which could alternatively be a fixed number of events) which
transits along a fragment of a score, measuring at each stage how
relevant musical information is organised. This way, certain critical
points arise, at which the form could be segmented, or perhaps find
tension or release points.

For the specific cases we have examined, analyzing a music score together
with its resulting series of ECG plots we can conclude that indeed
there are several relevant musical features that are somehow reflected
in one or more network metrics. Given a series of plots, there seems
to be a strong correlation between the frequency of certain peaks
or valleys, and the \emph{magnitude} of the change in the music. By
this, we mean there is a more drastic change in one or more music
features.

As ``control'' examples we have chosen some representative works
of European classical, though it is not our intention to restrict
ourselves to that repertoire. In fact, we are more interested in applying
the present data analysis methods to music scores outside the Western
corpus of the so-called common practice period. Yet we believe it
is pertinent to contrast our proposal with traditional tonal analysis.

Let us look closely at a simple example: the well-known Minuet in
G Major from the book for Anna Magdalena Bach. This piece shows a
stereotypical minuet form in a major key: thirty two bars, with the
first section englobing bars $1\text{\textendash}16$, a phrase in
the dominant in bars $17-24$, and the repeat of the second phrase
of the first section.
\begin{example}
\begin{figure}
\begin{centering}
\includegraphics[scale=0.5]{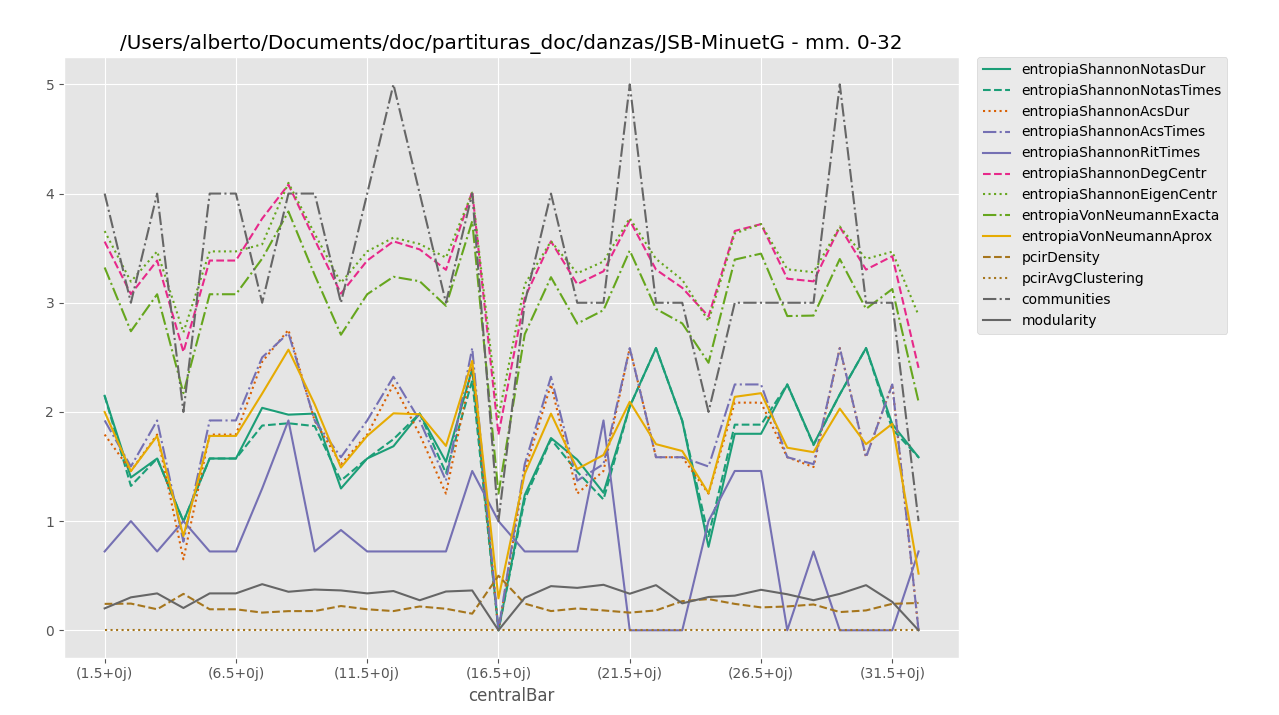}
\par\end{centering}

\protect\caption{{\footnotesize{}\label{fig:ECG-JSB-MinuetG_prueba_1c_sin_empalmar}ECGs
for metrics from the p-c-r graph for J. S. Bach's Minuet in G from
the Notebook for Anna Magdalena, taking time windows of length $1$
measure, without overlapping}.}
\end{figure}

\begin{figure}
\begin{centering}
\includegraphics[scale=0.5]{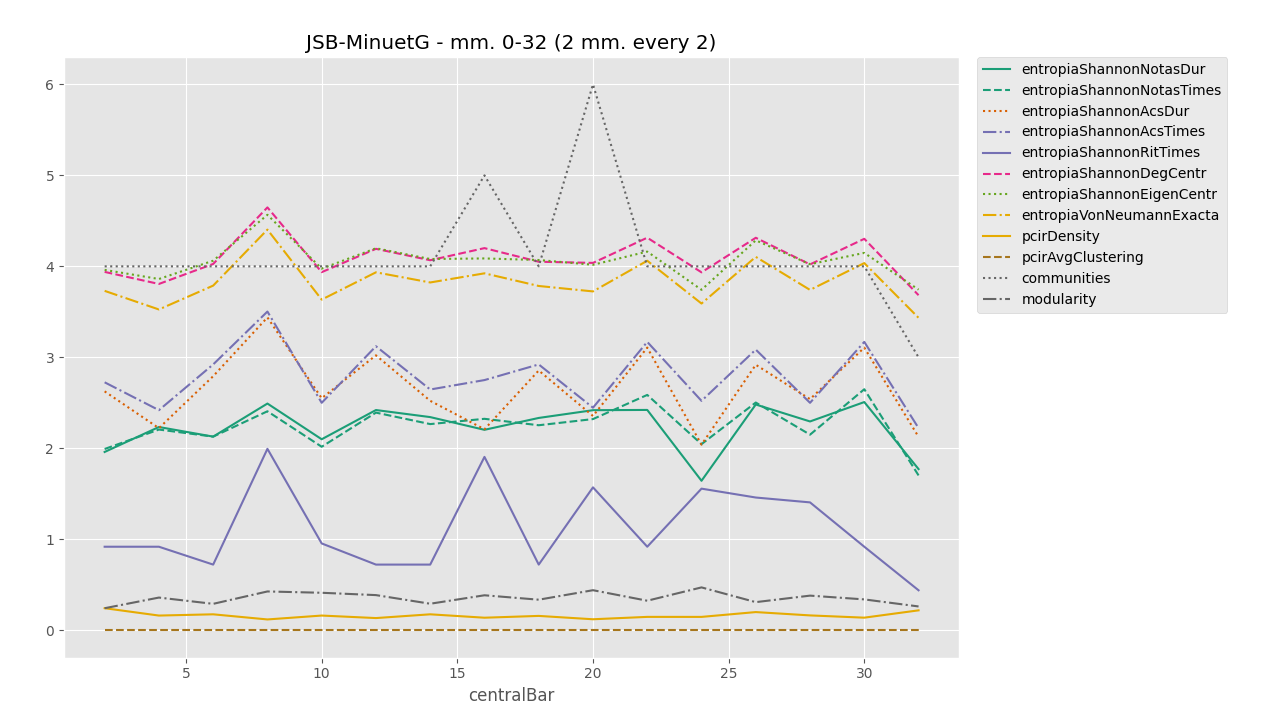}
\par\end{centering}

\protect\caption{{\footnotesize{}\label{fig:ECG-JSB-MinuetG_prueba_2cc_sin_empalmar}ECGs
for metrics from the p-c-r graph for J. S. Bach's Minuet in G from
the Notebook for Anna Magdalena, taking time windows of length $2$
mm., without overlapping}.}
\end{figure}

\begin{figure}
\includegraphics[scale=0.5]{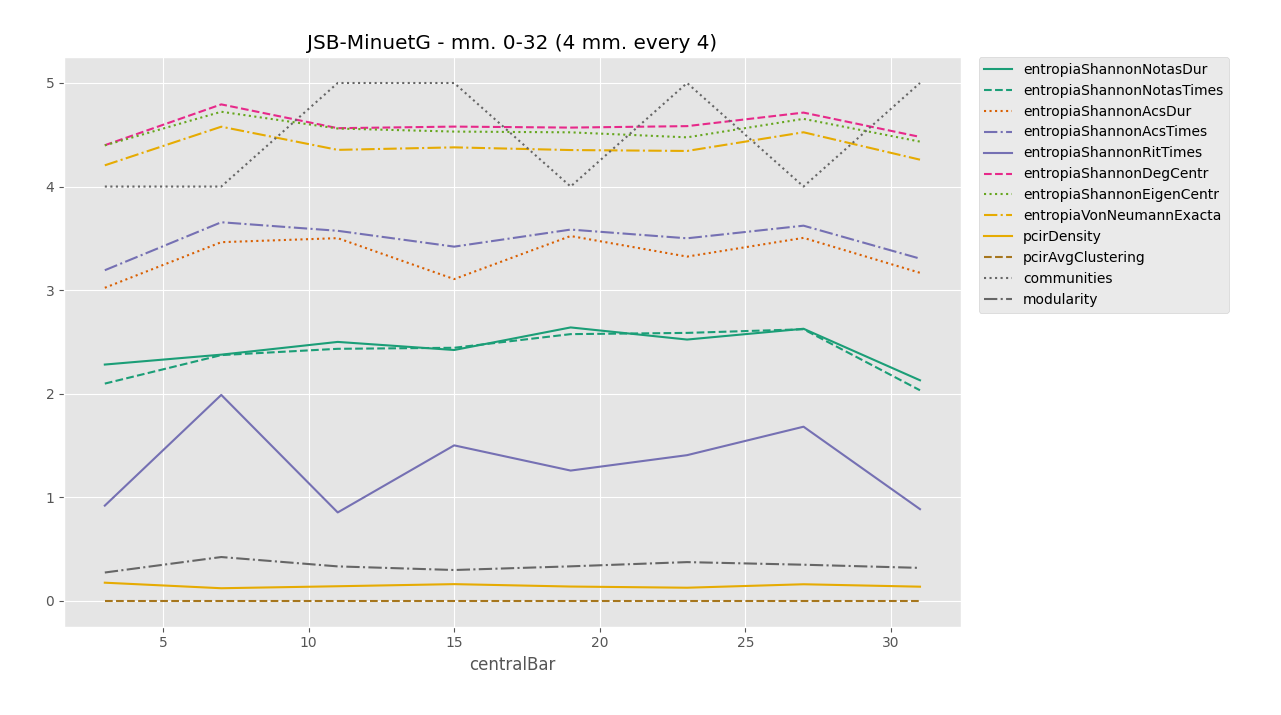}

\protect\caption{{\footnotesize{}\label{fig:ECG-JSB-MinuetG_prueba_4cc_sin_empalmar}ECGs
for metrics from the p-c-r graph for J. S. Bach's Minuet in G from
the Notebook for Anna Magdalena, taking time windows of length $4$
mm., without overlapping}.}
\end{figure}
\begin{figure}
\begin{centering}
\includegraphics[scale=0.5]{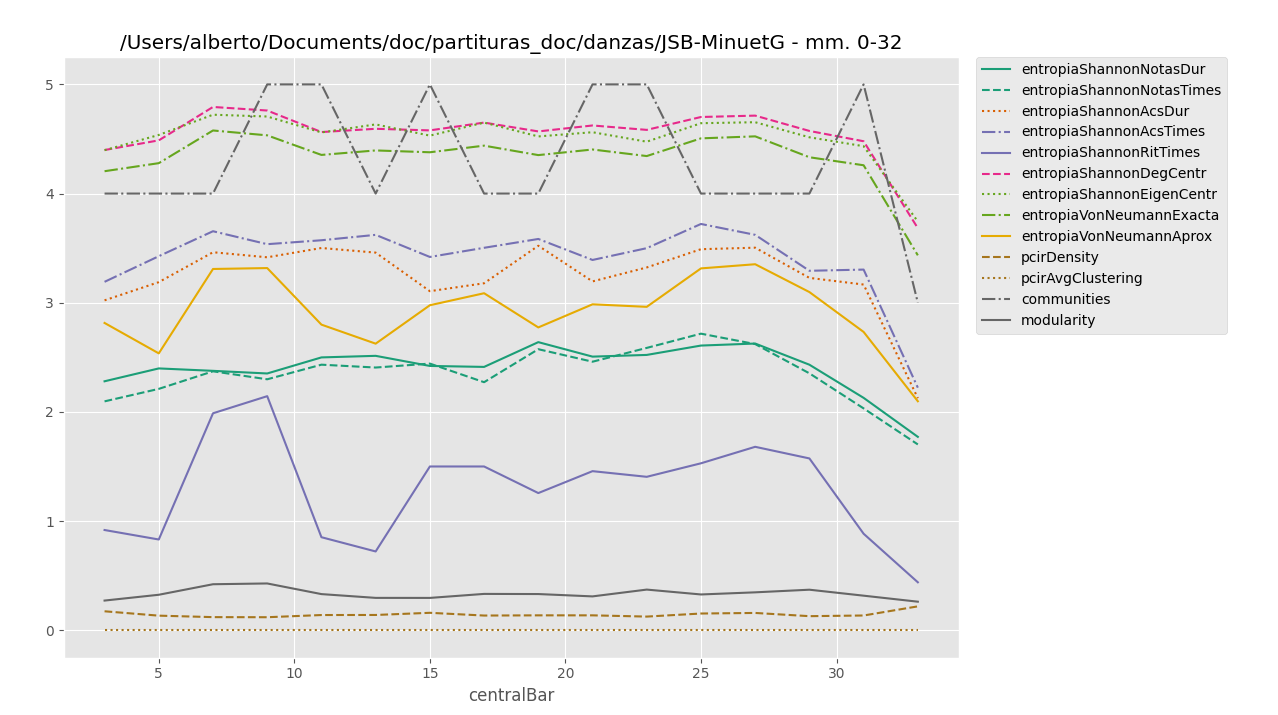}
\par\end{centering}

\protect\caption{{\footnotesize{}\label{fig:ECG-JSB-MinuetG_prueba_4cc_cada_2}ECGs
for metrics from the p-c-r graph for J. S. Bach's Minuet in G from
the Notebook for Anna Magdalena, taking time windows of length $4$
mm., every $2$ mm.}.}
\end{figure}
\begin{figure}
\begin{centering}
\includegraphics[scale=0.5]{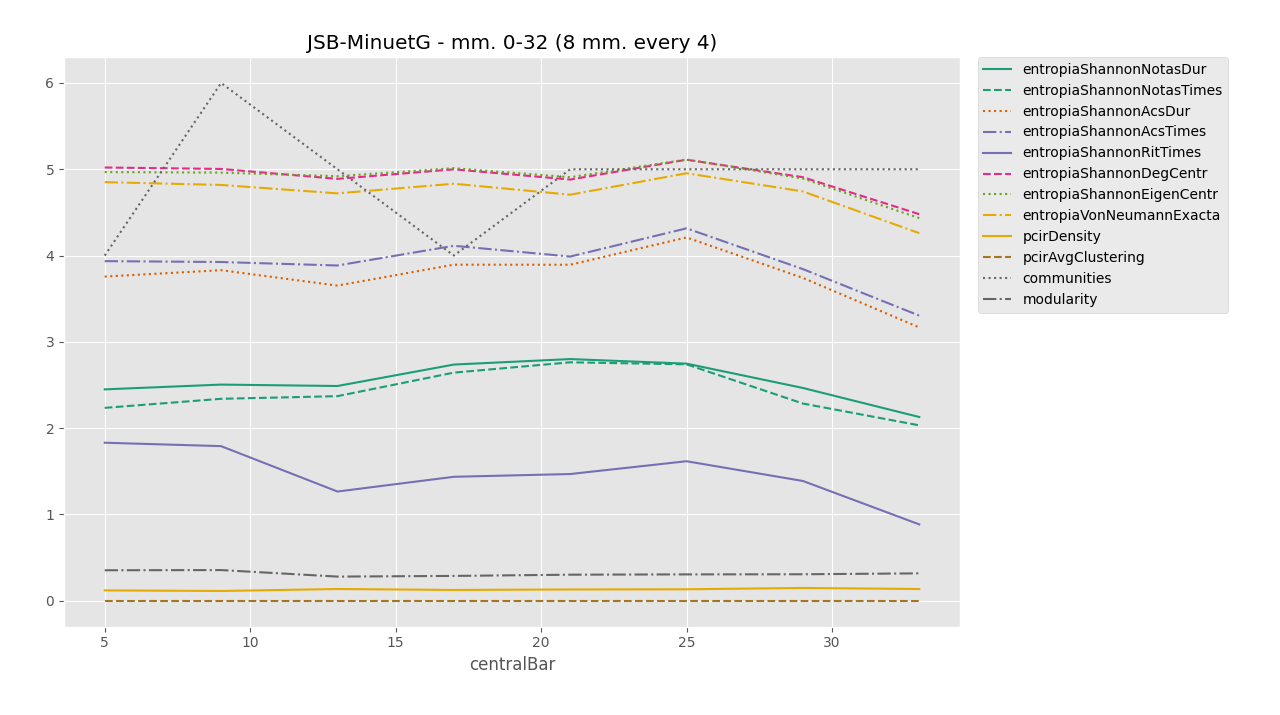}
\par\end{centering}

\protect\caption{{\footnotesize{}\label{fig:ECG-JSB-MinuetG_prueba_8cc_cada_4}ECGs
for metrics from the p-c-r graph for J. S. Bach's Minuet in G from
the Notebook for Anna Magdalena, taking time windows of length $8$
mm., every $4$ mm.}.}
\end{figure}

As we stated before, in the above plots we may observe several points
with recurrent presence of valleys or peaks. For example, around measures
$8$ and $16$, coincident with the end of the first and second phrases.
Around m. $21$ we also find what seem to be relevant peaks in the
plots, this time within a segment where we find a modulation to D
Major. Finally, we also notice a persistent peak around mm. $25\text{\textendash}27$,
coincident with the last phrase of the piece and a modulation back
to G Major (from D Major). 
\end{example}

\section{Conclusions and future work}

Besides looking deeper into emerging communities in music-defined
graphs and networks, we want to study their motifs and their possible
musical interpretation.

We will seek a deeper analysis of the time series obtained from a
larger score corpus. Additional to the use of dynamic time warping
to compare time series, we are interested in also using their visibility
graphs for such task.

It is projected to use hierarchical clustering algorithms for sample
classification.

We are currently seeking a mathematical criteria which could lead
to an automatic music form segmentation tool, based on entropy and
other graph properties. It would be interesting to also apply signal
processing methods, as in \citet{TonHarmTopDynNetworks} for the time
series defined by the metrics computed for the graphs and networks
hereby discussed.

Inquire in the behavior and utility of networks containing other music
parameters and relations: rhythm, dynamics, texture, etc. 

Inquire about the convenience or meaningfulness of keeping just some
of the metrics here considered (droping those which are too close
to each other.... or... are they subtle but meaningful for style distinction?).

\pagebreak{}

\bibliographystyle{plainnat}

\bibliography{/Users/alberto/Documents/doc_biblio/docBibTeX}

\begin{thebibliography}{15}
\providecommand{\natexlab}[1]{#1}
\providecommand{\url}[1]{\texttt{#1}}
\expandafter\ifx\csname urlstyle\endcsname\relax
  \providecommand{\doi}[1]{doi: #1}\else
  \providecommand{\doi}{doi: \begingroup \urlstyle{rm}\Url}\fi

\bibitem[Brown and George(2023)]{GrModPopMus}
Jason~I Brown and Ian George.
\newblock Modulation graphs in popular music.
\newblock \emph{arXiv preprint arXiv:2306.13691}, 2023.

\bibitem[Buongiorno~Nardelli(2023)]{TonHarmTopDynNetworks}
Marco Buongiorno~Nardelli.
\newblock Tonal harmony and the topology of dynamical score networks.
\newblock \emph{Journal of Mathematics and Music}, 17\penalty0 (2):\penalty0
  198--212, 2023.

\bibitem[Clauset et~al.(2004)Clauset, Newman, and
  Moore]{CommunitiesGreedyModularity}
Aaron Clauset, M.~E.~J. Newman, and Cristopher Moore.
\newblock Finding community structure in very large networks.
\newblock \emph{Physical Review E}, 70\penalty0 (6), dec 2004.
\newblock \doi{10.1103/physreve.70.066111}.
\newblock URL \url{https://doi.org/10.1103%2Fphysreve.70.066111}.

\bibitem[Cuthbert and Ariza(2010)]{Music21Library}
Michael~Scott Cuthbert and Christopher Ariza.
\newblock music21: A toolkit for computer-aided musicology and symbolic music
  data.
\newblock 2010.

\bibitem[Dehmer and Mowshowitz(2011)]{EntropyGraphsHist}
Matthias Dehmer and Abbe Mowshowitz.
\newblock A history of graph entropy measures.
\newblock \emph{Information Sciences}, 181\penalty0 (1):\penalty0 57--78, 2011.

\bibitem[Forte and Gilbert(1982)]{SchenkerianForte}
A.~Forte and S.E. Gilbert.
\newblock \emph{Introduction to Schenkerian Analysis}.
\newblock Norton, 1982.

\bibitem[Hagberg et~al.(2008)Hagberg, Swart, and S~Chult]{ExploringNetX}
Aric Hagberg, Pieter Swart, and Daniel S~Chult.
\newblock Exploring network structure, dynamics, and function using networkx.
\newblock Technical report, Los Alamos National Lab.(LANL), Los Alamos, NM
  (United States), 2008.

\bibitem[Kenley and Cho(2011)]{ProteinGraphEntropy}
Edward~Casey Kenley and Young-Rae Cho.
\newblock Detecting protein complexes and functional modules from protein
  interaction networks: A graph entropy approach.
\newblock \emph{Proteomics}, 11\penalty0 (19):\penalty0 3835--3844, 2011.

\bibitem[Lerdahl and Jackendoff(1996)]{LerdahlJackendoffGTTM}
Fred Lerdahl and Ray~S Jackendoff.
\newblock \emph{A Generative Theory of Tonal Music, reissue, with a new
  preface}.
\newblock MIT press, 1996.

\bibitem[M{\"u}ller(2007)]{DynTimeWarp}
Meinard M{\"u}ller.
\newblock Dynamic time warping.
\newblock \emph{Information retrieval for music and motion}, pages 69--84,
  2007.

\bibitem[Newman(2018)]{NetworksNewman}
Mark Newman.
\newblock \emph{Networks}.
\newblock Oxford University Press, 2018.

\bibitem[Nielsen(2009)]{StatAnMusicCorpora}
Johan Sejr~Brinch Nielsen.
\newblock Statistical analysis of music corpora, 2009.

\bibitem[Ren(2014)]{ComplMusPatt}
Iris~Yuping Ren.
\newblock Complexity of musical patterns.
\newblock \emph{University of Warwick}, 2014.

\bibitem[Szeto and Wong(2006)]{GrPattMatchPostTon}
Wai~Man Szeto and Man~Hon Wong.
\newblock A graph-theoretical approach for pattern matching in post-tonal music
  analysis.
\newblock \emph{Journal of New Music Research}, 35\penalty0 (4):\penalty0
  307--321, 2006.

\bibitem[Walton(2010)]{GrThApprTonMod}
Adrian Walton.
\newblock A graph theoretic approach to tonal modulation.
\newblock \emph{Journal of Mathematics and Music}, 4\penalty0 (1):\penalty0
  45--56, 2010.

\end{thebibliography}
\addcontentsline{toc}{section}{References}

\pagebreak{}

\selectlanguage{american}%
\appendix

%
%
%
%

\end{document}